\documentclass[12pt,fleqn]{article}
\newcommand{\ol}{\setlength{\itemsep}{0pt.}\begin{enumerate}}
\newcommand{\eol}{\end{enumerate}\setlength{\itemsep}{-\parsep}}

\begin{document}
\bibliographystyle{plain}
\parindent 0cm
\parskip 0.2cm
\input{psfig}
\newcommand{\bbox}{\vrule height8pt width6pt depth1pt}
\title
{ Attacks on Symmetric  Quantum Coin-Tossing Protocols}
\author
{Boaz Leslau\thanks{\mbox{\em E-mail:boazl@cs.huji.ac.il}}
\\
Institute of Computer Science,\\
Hebrew University \\
Jerusalem 91904\\ 
Israel}
\date{}

\maketitle
\begin{abstract}
We suggest an attack on a symmetric non-ideal quantum coin-tossing protocol  suggested by Mayers Salvail  and  Chiba-Kohno. The analysis of the attack shows that the  protocol is insecure.
\end{abstract}     

\section
{Introduction}

The subject of two party quantum cryptography has had many twists and turns and until now security was proven only for very weak  tasks like quantum gambling \cite{Vai:gam} and cheat sensitive bit commitment \cite{Ken:che}, \cite{Aha:esc}, while many two party tasks such as bit commitment \cite{May:imp}, \cite{Loc:imp} , ideal coin tossing \cite{Loc:ide} and secure two party computations~\cite{Lo:com} were proven insecure.

Recently, however, Mayers,Salvail and Chiba-Kohno \cite{May:coin} suggested a protocol, called thereafter MSC(99). They claimed this protocol may achieve unconditionally secure non-ideal coin tossing.

 \textit{Coin tossing} is a task in which two remote distrustfull parties conventionally called Alice  and Bob  run a protocol that, in the case that both parties act honestly, has equal probability to give the result 1 or 0. The protocol has another result \textit{abort} which is not  obtained  when both parties act honestly (except with a possibly arbitrarily small probability). Therefore when  \textit{abort}   is obtained the interpretation of an honest player is that  the other party deviated from the protocol. A protocol that satisfies the above conditions is called \textit{ correct}.

Let $ p_{b} ($ where $ b=0,1 $) be the probability that the result $ b $ is obtained. A  \textit{  non-ideal exact coin tossing protocol} is called \textit{secure} if, in the case that one of the parties cheats, (i.e. deviates from the given protocol) the following condition: $ p_{b}\leq 1/2 $, is satisfied. A \textit{ non-ideal non-exact coin tossing protocol}   is called \textit{secure} if, in the case that one of the parties cheats, the following   requirement is satisfied:  $ p_{b}\leq\frac{1}{2}+\xi$, where $ \xi $ is a security parameter that can be made arbitrarily small. 

The idea behind MSC(99), which is a \textit{  non-ideal non-exact coin tossing protocol},  is to build a protocol where both parties have almost no  information, on the completely random output at first. As the protocol proceeds, they get in a  slow and almost symmetric way more and more information about the result , until both have full information on the output. Such a protocol is supposed to overcome the  generalized attacks used against quantum bit commitment since, although at first a cheater can change the result without being detected,  he has no information about the result and therefore any change would be useless. While at the end of the protocol, a cheater may have information about the protocol's result but it will be almost impossible for him to change the result undetected. In this letter we will analyze the information flow in MSC(99)  and suggest  an attack  on MSC(99) in which  the cheater acts honestly on the quantum level (keeps everything possible in a superposition) until he can obtain enough (but not to much) information about the protocol's result. He then makes  the measurement and causes the superposition to ``collapse''  according to the information he obtains and his desired result. This attack is a generalization of the standard attacks on quantum bit commitment and can be carried out by any participant as long as the protocol is symmetric. This attack on MSC(99) creates a  non negligible bias on the protocol's  output  that is independent of the parameters of the protocol and therefore shows that the protocol is insecure.

\section
{Distinguishability measures}
In what follows will shall use the following results from quantum information theory. Given a quantum system whose density matrix is, with equal probability, one of two possible density matrices $\rho_{0},\rho_{1} \in \mathcal{H}_{1}$ and asked to decide which one we were given,  the measurement that minimizes the error probability of the decision is a measurement of the observable $\frac{ \rho_{0}-\rho_{1}}{2} $ and the probability of error is: 
\begin{equation}
PE(\rho_{0},\rho_{1})=\frac{1}{2}-\frac{1}{4}tr|\rho_{0}-\rho_{1}|,  \label{eq:PE}
\end{equation}
 \cite{Hel:det}. Another important quantity is the Kolmogorov distance, which is defined in the following way:
\begin{equation}
 K(\rho_{0},\rho_{1})=\frac{1}{2}max\sum_{\mu}|tr(\rho_{0}E_{\mu})-tr(\rho_{1}E_{\mu})|  \label{eq:KOL}
\end{equation}
 where the  maximization is done over all possible POVM's. The connection between these two quantities is the following:
 \begin{equation}
 K(\rho_{0},\rho_{1})=\frac{1}{2} tr|\rho_{0}-\rho_{1}|.  \label{eq:kolpe}
\end{equation} This shows that the measurements that optimizes K and PE (which we call $ E_{K}$ ) are identical. Therefore  the Kolmogorov distance  quantifies the deviation from a random guess  giving it a well defined operational meaning, .

The transition probability of two density matrices $\rho_{0},\rho_{1}$ is defined as  \begin{equation}
P(\rho_{0},\rho_{1})=max|\langle\psi_{0}|\psi_{1}\rangle|^{2},  \label{eq:trans}
\end{equation}
 where $|\psi_{0}\rangle,|\psi_{1}\rangle \in \mathcal{H}_{1} \otimes \mathcal{H}_{2}$ are purifications of $\rho_{0},\rho_{1}$ and the maximization  runs over all possible purifications of $\rho_{0}, \rho_{1}$. Moreover,  we can fix the purification of one density matrix and do the maximization over all purifications of the other \cite{Joz:fid}. Since different purifications of a density matrix in $\mathcal{H}_{1}$ are related by unitary transformations on $\mathcal{H}_{2}$ the maximum can be obtained by maximizing over unitary transformations on $|\psi_{0}\rangle$ or $|\psi_{1}\rangle$ in $\mathcal{H}_{2}$ alone. 

Another important quantity is the fidelity of two density matrices which is defined by:
\begin{equation}
F(\rho_{0},\rho_{1})=min \sum_{\mu}\sqrt{tr(\rho_{0}E_{\mu})} \sqrt{tr(\rho_{1}E_{\mu})}  \label{eq=fid}
\end{equation}
where we minimize over all possible POVM's. It was shown in \cite{fca:tech} that the fidelity can be written explicitly as 
\begin{equation}
F(\rho_{0},\rho_{1})=tr\sqrt{\sqrt{\rho_{0}}\rho_{1}\sqrt{\rho_{0}}}
\end{equation}
and in \cite{Uhl:fid} it was proven that the fidelity is related to the transition probability  through: 
\begin{equation}
F(\rho_{0},\rho_{1})=\sqrt{P(\rho_{0},\rho_{1})}.  \label{eq:fidtrans}
\end{equation}

\section
{The suggested attack on MSC(99)}

Let us, following MSC(99), define the normalized qubit states $\psi(0)=c|0\rangle+s|1\rangle$ and $\psi(1)=c|0\rangle-s|1\rangle$ where $c,s $ are real numbers.  Let $\Phi(b_{j})=\otimes_{k=1}^{n}\psi(b_{j})$, where $ b_{j}\in\{0,1\} $. A two element POVM with results $ \{c_{j} ; \bot \} $ where $ c_{j} \in \{0,1 \} $, is defined in the following way  $ E_{c_{j}}=|\Phi(c_{j})\rangle\langle\Phi(c_{j})| $, $E_{c_{j}}^{\bot}=\mathbf{1}-E_{c_{j}} $. Whenever the result $\bot$ is obtained the result of the protocol is abort.    

We will now describe in detail an explicate attack on the MSC(99) protocol.
We assume w.l.o.g that the cheater, whom we call Bob*, wishes to create a bias towards $0$ and that Alice is acting honestly throughout the protocol. We assume for the sake of simplicity that when both participants act honestly the probability for $\bot$ is 0. In the parenthesis we shall describe what an honest Bob does according to MSC(99).
\begin{description}

\item[Step 1] For $j=1 , \ldots , m$ do: \\
 Alice chooses randomly a bit $a_{j}\in\{0,1\}$. \\
 Bob*  does nothing. \\
(Bob chooses a random bit $ b_{j}\in \{0,1\} $. )  \\
	
\item[Step 2] For $i=1 ,  \ldots , n$ do: For $j=1, \ldots , m$ do: \\
Alice uniformly picks at random a bit $c_{ij}$ and sends a pair of  qubits in the state $\psi(c_{ij})\otimes\psi(\bar{c_{ij}})$. \\
Bob* picks at random a bit $d_{ij}$ and prepares  the following entangled state
\begin{equation}
|\eta\rangle=\frac{1}{\sqrt{2^m}}\sum_{b} ( \bigotimes_{j=1}^{m}\bigotimes_{i=1}^{n}\psi(b_{j}\oplus d_{ij})\psi(\bar{b_{j}}\oplus d_{ij}) ) |b\rangle_{B}
\end{equation}
where the $ \{ |b \rangle \}_{b=1}^{2^{m}}$ are $2^{m}$  orthogonal vectors in the Hilbert space $\mathcal{H}_{B}$, $| b\rangle=\otimes_{j=1}^{m}|b_{j}\rangle $.\\
 Bob* sends Alice the state $\rho=tr_{B}|\eta\rangle\langle\eta| $.\\
(Bob uniformly picks a random bit $d_{ij}$ and sends a pair of  qubits in the product state $ \psi(d_{ij})\psi(\bar{ d_{ij}})$.)  \\

\item[Step 3] For $i=1,\ldots , n $ do:For $j=1  , \ldots , m$ do:\\
Alice announces $e_{ij}=a_{j}\oplus c_{ij}$ and Bob* returns the second qubit at position $(i,j)$ if $e_{ij}=0$ and the first qubit otherwise. The Hilbert space of the qubits Bob* sends back will be called $\mathcal{H}_{ABA}$, and the qubits that Bob* keeps belong to the Hilbert space $\mathcal{H}_{AB}$.\\
 Bob* announces $f_{ij}=d_{ij}$ and Alice returns the second qubit at position $(i,j)$ if $f_{ij}=0$ and the first  qubit otherwise. The Hilbert space of the qubits Alice returns is $\mathcal{H}_{BAB}$. The Hilbert space of the qubits Alice keeps is $\mathcal{H}_{BA}$.\\
 At this stage, the  qubits in $\mathcal{H}_{AB}$, for every $j$ are in the state $\Phi(a_{j})$. The qubits in $\mathcal{H}_{ABA}$ are in the state $\Phi(\bar{a_{j}})$. 
The $mn$ qubits in $\mathcal{H}_{BAB}$ are entangled with the $mn$ qubits in $\mathcal{H}_{BA}$  and with the register in $\mathcal{H}_{B}$. Let us define: $|\Psi^{l}_{b}\rangle_{A,B}=\bigotimes_{j=l}^{m}\Phi(b_{j})_{A}\Phi(\bar{b_{j}})_{B}$ then the entangled state shared by Alice and Bob* is:
\begin{equation}
|\eta'\rangle=\frac{1}{\sqrt{2^m}}\sum_{b}|\Psi_{b}^{1}\rangle_{BA,BAB} |b\rangle_{B}.
\end{equation} 
(Bob announces $f_{ij}=b_{j}\oplus d_{ij}$ and Alice returns the second qubit at position $(i,j) $ if $ f_{ij}=0 $ and the first qubit otherwise.)

\item[Step 4A] 
For $j=1 , \ldots , l-1 $ do: \\
Alice announces $a_{j}$, Bob* executes the POVM  $(E_{a_{j}}, E_{a_{j}}^{\bot})$  on $\Phi(a_{j})$ notes the outcome $\tilde{a}_{j}$ and if $\tilde{a}_{j}=\bot$ the protocol aborts.\\ 
Bob* measures the following two element POVM  
$D_{0}^{j}=|0\rangle\langle 0|$ 
and 
$D_{1}^{j}=|1\rangle\langle 1|$ 
in $\mathcal{H}_{B}$ on the j-th qubit. If he gets the result $D_{0}^{j}$ he announces $ b_{j}=0 $ and if he receives the result $D_{1}^{j}$ he announces $ b_{j}=1 $. Alice executes the POVM $(E_{b_{j}},E_{b_{j}}^{\bot})$ on $\Phi(b_{j})$ if the outcome is $\tilde{b_{j}}=\bot $ the protocol aborts. \\
 At this stage the entangled  state in $ \mathcal{H}_{BA} \bigotimes \mathcal{H}_{BAB} \bigotimes \mathcal{H}_{B} $ is:
\begin{equation}
|\eta''\rangle=\frac{1}{\sqrt{2^{m-j}}}\sum_{b'}|\Psi_{b'}^{j+1}\rangle_{BA,BAB} |b'\rangle_{B}
\end{equation} 
where $ b'\in \{0,1\}^{m-j}$. \\ 
(Bob announces $ b_{j} $, Alice executes the POVM $(E_{b_{j}},E_{b_{j}}^{\bot})$ on 
$\Phi(b_{j})$, if the outcome is $\tilde{b_{j}}=\bot $ the protocol aborts.) \\

\item[Step 4B]
 For $ j=l $ do: \\
 Alice announces $ a_{l} $ Bob* executes the above POVM on $\Phi(a_{l})$ and if $ \tilde{a_{l}}=\bot$ the protocol aborts.\\
Let:
\begin{equation}
\rho_{C}^{0}(k)=\frac{1}{2^{m-k}}\sum_{\{a^{k}|\oplus_{j=k}^{m}a_{j}=0\}}(\bigotimes_{j=k}^{m}\Phi(a_{j})\Phi^{\dagger}(a_{j}))
\end{equation}
\begin{equation}
\rho_{C}^{1}(k)=\frac{1}{2^{m-k}}\sum_{\{a^{k}|\oplus_{j=k}^{m}a_{j}=1\}}(\bigotimes_{j=k}^{m}\Phi(a_{j})\Phi^{\dagger}(a_{j}))
\end{equation}
where $ C\in\{A,B\} $ 	and  $a^{k}\in\{0,1\}^{m-k}$. Bob* measures the remaining $(m-(l+1))n $ qubits in $\mathcal{H}_{AB}$ with the POVM that  distinguishes maximally between $ \rho_{B}^{0}(l+1) $ and $ \rho_{B}^{1}(l+1)$ and obtains the result $\tilde{A}''\in\{0,1\}$.

Bob* then measures the following POVM on $\mathcal{H}_{B}$ 
\begin{equation}
F_{0}=\sum_{\{b''|\oplus_{j=l}^{m}b_{j}=0\} }|b''\rangle\langle b''|
\end{equation}
 and 
\begin{equation}
F_{1}=\sum_{\{b''|\oplus_{j=l}^{m}b_{j}=1\}} |b''\rangle\langle b''|
\end{equation}
where $b''\in\{0,1\}^{m-l+1}$, and obtains the result $F\in \{0,1\}$. Bob* calculates the following expression $\tilde{X}= \oplus_{j=1}^{l}\tilde{a_{j}}\oplus_{j=1}^{l-1}b_{j}\oplus\tilde{A''} \oplus F$. 
 At this stage the state entangled between Bob* and Alice is the following:
\begin{equation}
|\eta''(F)\rangle=\frac{1}{\sqrt{2^{m-l}}}\sum_{\{b''|\sum_{j=l}^{m}b_{j}=F\}} |\Psi_{b''}^{l}\rangle_{BA,BAB}|b''\rangle_{B}
\end{equation}
and the state in $\mathcal{H}_{BA}$ is $\rho_{A}^{F}(l)=Tr_{B,BAB}(|\eta''(F)\rangle \langle \eta''(F)|)$. \\
If  $\tilde{X}=0$, Bob* measures the POVM with the elements 
\begin{equation}
 B_{b''}^{F}=|b''\rangle\langle b''|,
\end{equation}
where $ b''=\{b''|\oplus_{j=l}^{m}b''_{j}=F\} $. 
 Bob* obtains the result $b''$ and announces $b''_{l}$.\\
If  $\tilde{X}=1$ Bob*  does the following:

We know from (\ref{eq:trans}) that:
\begin{equation}
 u=P(\rho_{B}^{1}(l),\rho_{B}^{0}(l))=max\mid\langle\eta''(\bar{F})|U_{B}\eta''(F)\rangle \mid^{2}
\end{equation}
where maximization is done over all unitary transformations $U_{B}$ in $\mathcal{H}_{B}$. Let $U^{*}_{B}$ be the transformation that achieves  the maximum, then  $U^{*}_{B}|\eta''(F)\rangle$ can be written in the following way:
\begin{eqnarray}
U^{*}_{B}|\eta''(F)\rangle & = &  \sqrt{u}e^{i\theta}|\eta''(\bar{F})\rangle+\sqrt{1-u}e^{i\varphi}|\overline{\eta''(\bar{F})}\rangle \nonumber \\ 
& = & \sqrt{\frac{u}{2^{m-l}}}e^{i\theta}\sum_{\{b''|\oplus_{j=l}^{m} b_{j}=\bar{F}\}}|\Psi_{b''}^{l}\rangle_{BA,BAB}|b''\rangle_{B}  \nonumber \\
& & +\sqrt{1-u}e^{i\varphi}|\overline{\eta''(\bar{F})} \rangle. 
\end{eqnarray}
Bob* applies the transformation $ U^{*}_{B} $ on $ \mathcal{H}_{B} $, and  measures a  POVM with the elements $ B_{\hat{b''}} $, where $ \hat{b''}=\{ b''| \oplus_{j=l}^{m}b''_{j}=\bar{F} \} $. When Bob obtains the result $\hat{b''}=\hat{b''_{l}}\ldots \hat{b''_{m}}$  he announces $\hat{b''_{l}}$. Otherwise he chooses randomly a $\hat{b''}$ where $\oplus_{k=l}^{m} \hat{b''_{k}}=\bar{F}.$\\
 Alice executes the POVM $(E_{b_{l}},E_{b_{l}}^{\bot})$ on the $ l$-th qubit if the outcome  is $\tilde{b_{l}}=\bot $ the protocol aborts.\\
(Bob announces $ b_{l} $. Alice executes the POVM $(E_{b_{l}},E_{b_{l}}^{\bot})$ on $\Phi(b_{l})$ if the outcome is $\tilde{b_{l}}=\bot $ the protocol aborts.)  

\item[Step 4C] 
For $j=l+1, \ldots , m$ do: \\
 Alice announces $a_{j}$ Bob* does nothing. \\
 Bob* announces $b_{j}''$ if $\tilde{X}=0$ or $\hat{b_{j}''}$ if $\tilde{X}=1 $.  Alice executes the POVM $(E_{b_{j}},E_{b_{j}}^{\bot})$ on the $j$-th qubit, if the outcome is $\tilde{b_{j}}=\bot $ the protocol aborts. 

The result of the coin tossing  will be $ X=\oplus_{j=1}^{l}\tilde{a_{j}}\oplus_{j=l+1}^{m}a_{j}\oplus_{j=1}^{m}\tilde{b_{j}}=\oplus_{j=1}^{l}\tilde{a_{j}}\oplus A''\oplus_{j=1}^{l-1}\tilde{b_{j}}\oplus\tilde{B} $.\\
(Bob announces $ b_{j} $. Alice executes the POVM $(E_{b_{j}},E_{b_{j}}^{\bot})$ on $\Phi(b_{j})$ if the outcome is $\tilde{b_{j}}=\bot $ the protocol aborts.)

\item [Step 5]
For $j=1 \ldots m$ do: \\
Alice measures the state $\Phi(\bar{a_{j}})$ returned by Bob* at position $j$ with the POVM $(E_{\bar{a_{j}}}, E_{\bar{a_{j}}}^{\bot})$ and if the outcome is $\bot$ the protocol aborts.\\
Bob*  measures the state $\Phi(\bar{b_{j}})$ returned by Alice at position $j$ with the POVM $(E_{\bar{b_{j}}},E_{\bar{b_{j}}}^{\bot})$ and if the outcome is $\bot$ the protocol aborts.\\
(Bob measures the state $\Phi(\bar{b_{j}})$ returned by Alice at position $j$ with the POVM $(E_{\bar{b_{j}}},E_{\bar{b_{j}}}^{\bot})$ and if the outcome is $\bot$ the protocol aborts.)
\end{description}

\section{Analysis of the attack}

Let us calculate the bias Bob* created on the distribution of the correct results $(0,1)$ of the protocol . In order to obtain this bias we must calculate what is the probability that the result of  the coin tossing is $0$ and that the protocol is found to be correct. This will be in our case  $p(X=0)$ because for $ z \in \{ 0,1,\bot \}$ we have that $z \oplus \bot= \bot $.

 The probability that the result of the coin tossing is $0$ or $  \bot $, is the same as  the probability that $\tilde{A''}= A''$, because whenever Bob* obtains an error  at step 4B   the protocol's result is 1 or $ \bot $. 
 This probability is:
\begin{equation}
p(\tilde{A''}=A'')=1-PE(\rho^{0}_{B}(l+1),\rho_{B}^{1}(l+1))=1-PE_{B}^{l+1}(\rho_{0},\rho_{1}).
\end{equation}

Since we assume Alice is acting honestly,and until step 4B Bob* always gives Alice results that coincide with what Bob would have told her, ( With  probability one  the protocol does not abort until that step, since we assumed that the honest protocol aborts with probability zero.) therefore  the probability that the protocol succeeds is:$\; p(X \neq \bot )=p(\tilde{B} \neq \bot )$.  In the case  Bob* obtains in step 4B  $\tilde{X}=0$:
\begin{eqnarray}
p(B\neq\bot) & = & \sum_{b''} p(\tilde{b''}=b'')\nonumber \\
& =  &\sum_{b''}\frac{1}{2^{m-l}}|\langle b'' | \langle\Phi(\tilde{b'')}|\eta''(F)\rangle|^{2}=1.
\end{eqnarray}
If  the result Bob* obtains in step 4B is $\tilde{X}=1$ the probability  that the protocol succeeds will be :
\begin{eqnarray}
p(B\neq\bot) & = &\sum_{\hat{b''}}p(\tilde{b''}=\hat{b''}) \nonumber \\
 &  =  & \sum_{\hat{b''}}\frac{1}{2^{m-l}}|\langle \hat{b''} | \langle\Phi(\tilde{b'')}|U^{*}_{B}(|\eta''(F)\rangle)|^{2} \\
& \geq & u= P(\rho_{A}^{0}(l),\rho_{A}^{1}(l))=(F^l(\rho_{0},\rho_{1}))^2.
\end{eqnarray}

Bob obtains the results $0,1$ for $ A'' $ with probability  $ \frac{1}{2}$.  Therefore we obtain the following expression for the probability Bob* cheats successfully:
\begin{eqnarray}
p(X=0) & \geq  & \sum_{A''}p(A'')p(\tilde{A''}=A'')p(B\neq\bot) \nonumber \\ 
& = & \frac{1}{2}(1-PE^{l+1}(\rho_{0},\rho_{1})(1+(F^{l}(\rho_{0},\rho_{1}))^2).  \label{eq:bias}
\end{eqnarray}

The next step is an explicit calculation of (\ref{eq:bias}). The  probability of error in guessing the parity bit of a string of q bits with parameters c,s was 
found in \cite{May:coin} to be:
\begin{equation}
PE^{q}(\rho_{0},\rho_{1})= \frac{1-(2cs)^{q}}{2}. \label{eq:err}
\end{equation}
The fidelity between the two density matrices was calculated in \cite{Fuc:cry} with the result:
\begin {equation}
F^q(\rho_{0},\rho_{1})=\sum_{k=0}^{[\frac{m-q}{2}]} 
{m-q  \choose k} |c^{2(m-q-k)}s^{2k}-c^{2k}s^{2(m-q-k)}|.  \label{eq:fidel}
\end{equation}

Unlike the case of \cite{May:coin} and \cite{Fuc:cry}, in MSC(99)   the quantum states that represent the two possible values of the bit $|\Phi(0)\rangle,|\Phi(1)\rangle$  belong to $ \mathcal{H}^{2^{n}}$. However they span a two dimensional Hilbert space $\mathcal{H'}$ therefore we can define $c'^2-s'^2=(c^2-s^2)^n $ and write $|\Phi(0)\rangle=c'|0'\rangle+s'|1'\rangle , |\Phi(1)\rangle=c'|0'\rangle-s'|1'\rangle $ where $|0'\rangle,|1'\rangle$ are two orthogonal vectors in $ \mathcal{H'}$. Therefore we can use the results of \cite{May:coin} and \cite{Fuc:cry} to write that:
\begin{eqnarray}
p(X=0) &\geq &\frac{1}{4}(1+(2c's')^{m-(l+1)}) \nonumber \\
&  &  (1+(\sum_{k=0}^{[\frac{m-l}{2}]}
{m-l\choose k}|c'^{2(m-l-k)}s'^{2k}-c'^{2k}
s'^{2(m-l-k)}|)^2).
\end{eqnarray}
A good  approximation (for large $m-l$) of this expression can be obtained in the following way: Let us define $ t=s'^2$, one easily recognizes that (\ref{eq:fidel}) is the statistical overlap of two binomial distributions with mean $t$ and $1-t$. Since, the transformation $ k \rightarrow m-k $, transforms one distribution to the other we can write
\begin{equation}
F^{l}(\rho_{0},\rho_{1})=2\sum_{k=0}^{[\frac{m-l}{2}]} {m-l \choose k} t^{k}(1-t)^{m-l-k}-1.  
\end{equation}
Using the following theorem \cite{Ren:prob}
\begin{equation}
 \sum_{k \leq nt+\alpha \sqrt{t(1-t)n} }  {n \choose k} t^{k}(1-t)^{n-k}=\frac{1}{2 \pi} \int_{-\infty}^{\alpha}e^{\frac{-x^2}{2}}dx+O (\frac{1}{\sqrt{n}})
\end{equation}
 we have that 
\begin{equation}
F^l(\rho_{0},\rho_{1})  \approx \sqrt{\frac{2}{\pi}} \int_{-\infty}^{\alpha}e^{\frac{-x^2}{2}}dx-1
\end{equation}
where $\alpha=\sqrt{\frac{(m-l)(1-2t)^2}{4(1-t)t}}$.  Therefore $t=\frac{1}{2}- \frac{\alpha}{2\sqrt{m-l+\alpha^{2}}}$ if $ t \leq 1-t$, and $t=\frac{1}{2}+ \frac{\alpha}{2\sqrt{m-l+\alpha^{2}}}$ if $t \geq 1-t$.
This leads to the following bound for the fidelity
\begin{equation}
F^l(\rho_{0},\rho_{1}) \approx  Erf(\frac{\alpha}{\sqrt{2}}).
\end{equation}
For large $m-l$ , we can simplify the probability that Bob* guessed correctly the result of the protocol at step $l+1$ in the following way:
\begin{eqnarray} 
1-PE^{l+1}(\rho_{0},\rho_{1}) \geq 1-PE^{l}(\rho_{0},\rho_{1}) & = &\frac{1+(4t(1-t))^{\frac{m-l}{2}}}{2} \nonumber \\
& =& \frac{1+(1-\frac{\alpha^2}{m-l+\alpha^2})^{\frac{m-l}{2}}}{2} \nonumber \\
 &  \approx & \frac{1+e^{\frac{-\alpha^2}{2}4t(1-t)}}{2}  \\
& \geq & \frac{1+e^{\frac{-\alpha^2}{2}}}{2}.
\end{eqnarray}
Therefore the probability the protocol's result is 0 is:
\begin{eqnarray}
P(X=0) & \approx & \frac{(1+e^{\frac{-\alpha^2}{2}})(1+(Erf(\frac{\alpha}{\sqrt{2}}))^2)}{4} \nonumber \\
& = & \frac{(1+K)(1+(Erf(\sqrt{-\ln K}))^2)}{4}.  \label{eq:fin}
\end{eqnarray}

In MSC(99) (and in any reasonable coin-tossing protocol) K  varies between a very small number in the beginning of the protocol, ( $(1-(c^2-s^2)^{2\log{M}})^M $ in our case) to very close to one at the end of the protocol, ($ 1-(c^2-s^2)^{2\log{M}}$ in our case)since K quantifies the participant's knowledge about the protocol's result. Figure 1 (below) shows that Bob*'s bias  is larger then 1/2 for all $0 < K < 1$ and that it reaches it's maximum  0.09195 when $K=0.510964$.

\begin{picture}(600,120)
\put(20,15){\psfig{figure=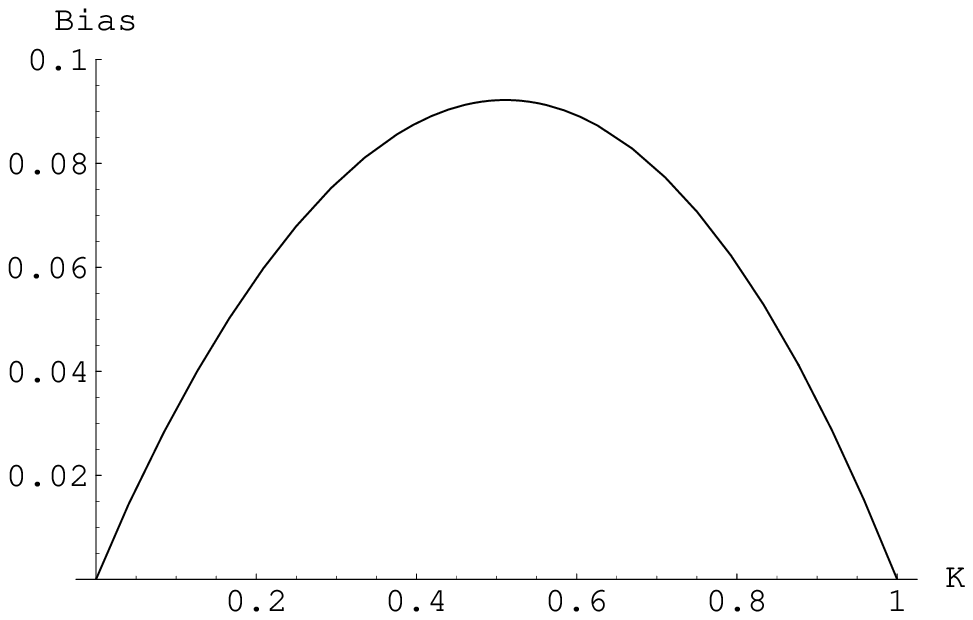,height=3.5cm}}
\put(0,0){\makebox(200,120)[b]{Fig 1. Bob*'s bias during the protocol}}
\end{picture}

These results  do not depend on the parameters (c,s,n,m) of the protocol, (as long as m is large enough and c not to close to $1/2$ which are necessary for the protocol to be secure and correct.) the bias created by  Bob* is  intrinsic to the  information fidelity tradeoff of the parity bit problem. In addition it gives us a constructive answer when should  a cheater attack in order to the obtain maximum bias for our attack. If we use the parameters suggested in MSC(99) ($c^2-s^2=\cos{\frac{\pi}{9}}$, $n=\log{m}$ ,m) a simple calculation shows that the optimal attack is when  $l=\frac{log{0.510964}}{log{\sqrt{(1-(\cos{\frac{\pi}{9}})^{2log{m}}}}} $ of the m bits have not been revealed.

It is important to note that we have not claimed  that this attack achieves maximum bias. The attack we suggested is oblivious to the exact structure of the density matrices $\rho_{C}^{0}(k),  \rho_{C}^{1}(k) $, therefore it is possible that a  cheater  can create  a larger bias on the protocol's result by  taking the structure of these density matrices into consideration.

\section{Conclusion}
The results obtained above do not completely obliterate the possibility of achieving a secure quantum coin tossing using a slow symmetric protocol such as MSC(99). The attack we suggested on such protocols shows, that protocols using this method must take great care of the tradeoff (such as in (\ref{eq:bias})) between the information a participant can obtain about the protocol's result  and the fidelity between the density matrices representing the protocol's possible results,  during the course of the protocol. 

\section{Acknowledgements}
I would like to thank Michael Ben-Or for many insightful discussions, and Andris Ambainis, Dominic Mayers  Louis Salvail and Yuki Tokunaga for their comments. This work was supported by the US-Israel Binational Science Foundation grant 9800229, by an EU-FPS QAIP research grant and by the Leibniz center.


\begin{thebibliography}{99}
\bibitem{Vai:gam} L. Goldenberg, L. Vaidman and S. Weisner, Phys. Rev. Lett. 82, 3356 (1999).
\bibitem{Ken:che}  L. Hardy and A. Kent, quant-ph/9911043.
\bibitem{Aha:esc} D. Aharonov, A. Ta-shma, U. Vazirani and A. Yao, Proc. of STOC00, p.705.
\bibitem{May:imp} D. Mayers, Phys. Rev. Lett. 78,  3414 (1997).
\bibitem{Loc:imp} H.-K. Lo and H. Chau, Phys. Rev Lett. 78, 3410 (1997).
\bibitem{Loc:ide} H.-K. Lo and H. Chau, Physica D. 120, 177 (1998).
\bibitem{Lo:com} H.-K. Lo, Phys. Rev. A 56 1154 (1997).  
\bibitem{May:coin} D. Mayers, L. Salvail, and Y. Chiba-Kohno, quant-ph/990478.
\bibitem{Hel:det} C. W. Helstorm,\emph{ Quantum Detection and Estimation Theory} (Academic press, New-York, 1976).
\bibitem{Joz:fid} R. Jozsa, J. Mod. Opt. 41, 2315 (1994).    
\bibitem{fca:tech} C. A. Fuchs, C. Caves, Open Systems and Information Dynamics, 3,3, 345 (1995).
\bibitem{Uhl:fid} A. Uhlmann, Rep. Math. Phys. 9, 273 (1976)
\bibitem{Fuc:cry} C. Fuchs, J. van de Graaf, quant-ph/9712042.
\bibitem{Ren:prob} A. Renyi, \emph{ Foundations of Probability} 
 (Holden-Day, San Francisco, 1970) p.210
\end{thebibliography}
\end{document}